\documentclass[12pt,preprint]{aastex}

\shorttitle{eruptive and confined solar flares} \shortauthors{Li et
al.}

\begin{document}

\title{A New Magnetic Parameter of Active Regions Distinguishing Large Eruptive and Confined Solar Flares}

\author{Ting Li\altaffilmark{1,2}, Xudong Sun\altaffilmark{3}, Yijun Hou\altaffilmark{1,2}, Anqin Chen\altaffilmark{4}, Shuhong Yang\altaffilmark{1,2} \& Jun
Zhang\altaffilmark{5}}

\altaffiltext{1}{CAS Key Laboratory of Solar Activity, National
Astronomical Observatories, Chinese Academy of Sciences, Beijing
100101, China; liting@nao.cas.cn} \altaffiltext{2}{School of
Astronomy and Space Science, University of Chinese Academy of
Sciences, Beijing 100049, China} \altaffiltext{3}{Institute for Astronomy, University of Hawai`i at M\={a}noa, Pukalani, HI 96768, USA}
\altaffiltext{4}{Key Laboratory of
Space Weather, National Center for Space Weather, China
Meteorological Administration, Beijing 100081, China}
\altaffiltext{5}{School of Physics and Optoelectronics engineering, Anhui University, Hefei 230601, China}

\begin{abstract}

With the aim of investigating how the magnetic field in solar active regions (ARs) controls flare activity, i.e., whether a confined
or eruptive flare occurs, we analyze 106 flares of Geostationary Operational Environmental
Satellite (GOES) class $\geq$M1.0 during 2010$-$2019. We calculate mean characteristic twist parameters $\alpha$$_{FPIL}$ within the ``flaring polarity
inversion line" region and $\alpha$$_\mathrm{HFED}$ within the area of high photospheric magnetic free energy
density, which both provide measures of the nonpotentiality of AR core region. Magnetic twist is thought to be related to the driving force of electric current-driven
instabilities, such as the helical kink instability.
We also calculate total unsigned magnetic flux ($\Phi$$_\mathrm{AR}$) of ARs producing the flare, which
describes the strength of the background field confinement. By considering both the constraining effect of background
magnetic fields and the magnetic non-potentiality of ARs, we propose a new parameter $\alpha$/$\Phi$$_\mathrm{AR}$ to measure the probability for a large flare to be associated with a coronal mass ejection (CME).
We find that in about 90\% of eruptive flares, $\alpha$$_\mathrm{FPIL}$/$\Phi$$_\mathrm{AR}$ and $\alpha$$_\mathrm{HFED}$/$\Phi$$_\mathrm{AR}$ are beyond critical values (2.2$\times$$10^{-24}$
and 3.2$\times$$10^{-24}$ Mm$^{-1}$ Mx$^{-1}$),
whereas they are less than critical values in $\sim$ 80\% of
confined flares. This indicates that the new parameter $\alpha$/$\Phi$$_\mathrm{AR}$ is well able to distinguish eruptive flares from confined flares.
Our investigation suggests that the relative measure of magnetic nonpotentiality within the AR core over the restriction of the background field
largely controls the capability of ARs to produce eruptive flares.

\end{abstract}


\keywords{Sun: activity---Sun: coronal mass ejections (CMEs)---Sun:
flares}

\section{Introduction}

Solar flares and coronal mass ejections (CMEs) are the rapid release of
a huge amount of magnetic energy accumulated in the solar corona through magnetohydrodynamic
instabilities and magnetic reconnection. Solar flares are often, but not always, accompanied by CMEs.
We refer to flares with a CME as ``eruptive flares" and flares not associated with a CME as ``confined flares".
It is revealed that flare-CME association rate increases with the flare intensity (Andrews 2003; Yashiro et al. 2006).
Recently, Li et al. (2020, 2021) found that flare-CME association rate decreases with
total unsigned magnetic flux ($\Phi$$_\mathrm{AR}$) of active regions (ARs) producing the flare,
which provides a global parameter relating to the strength of the background field confinement.

Over the last two decades, significant progress has been made in understanding the physical factors
determining whether a flare event is associated with a CME or not. It is suggested that eruptive flares tend to
occur if the overlying background magnetic fields are weaker or more quickly decay with height
(T{\"o}r{\"o}k \& Kliem 2005; Wang \& Zhang 2007; Wang et al. 2017; Baumgartner et al. 2018; Amari et al. 2018; Jing et al. 2018; Duan et al. 2019). Moreover, the
magnetic non-potentiality of ARs is thought to be another important factor governing the eruptive character of solar
flares (Nindos \& Andrews 2004; Liu et al. 2016; Cui et al. 2018; Vasantharaju et al. 2018; Thalmann et al. 2019; Avallone \& Sun 2020; Gupta et al. 2021),
such as free magnetic energy, relative helicity, magnetic twists, etc. Statistical studies have
shown that CME productivity is correlated with the twist parameter $\alpha$ (Falconer et al. 2002, 2006), which characterizes
the degree to which the photospheric magnetic fields of an AR deviate from a potential field (Leka \& Skumanich 1999; Yang et al. 2012). Bobra \& Ilonidis (2016)
found that the twist parameter $\alpha$ and mean gradient of the horizontal field are two relatively high-performing features
in predicting CMEs based on machine-learning algorithms. However, the measurement of previous known parameters showed that there are high degrees of overlap
between confined and eruptive flares if only one factor (overlying confinement or magnetic non-potentiality) was considered in selecting the parameters (Nindos \& Andrews 2004; Wang et al. 2017; Vasantharaju et al. 2018).

Sun et al. (2015) suggested that AR eruptivity is related to the relative value
of magnetic nonpotentiality over the restriction of the background field. In this Letter, following the idea of Sun et al. (2015), we consider both the constraining effect of background
magnetic fields and the magnetic non-potentiality of ARs, and propose a new parameter $\alpha$/$\Phi$$_\mathrm{AR}$
to describe the eruptive character of a flare. By measuring the parameter of 106 large flares within the cores of 12 CME-active and 9
CME-quiet ARs, we find that the new parameter $\alpha$/$\Phi$$_\mathrm{AR}$ is well able to distinguish flares associated with CMEs from flares
that are not.
\section{Database Selection and Parameter Calculations}

We use a subset of 106 flare events\footnote{\url{https://doi.org/10.12149/101087}} $\geq$M1.0 (43 eruptive and 63
confined) from a large database of 322 M-class flares\footnote{\url{https://doi.org/10.12149/101030}} during
the period of June 2010 to June 2019 (Li et al.
2020). The subset is selected based on the characteristics of ARs, and the selected ARs must fulfill the following two selection criteria.
First, the ARs are flare-active and produced $\geq$3 M-class flares.
Second, the ARs can be unambiguously classified into CME-active and CME-quiet. We refer to an AR as a CME-active (CME-quiet) AR if the flares from it are all eruptive (confined) or only one exceptional flare is confined (eruptive).
Finally, a total of 21 ARs fulfill these selection criteria, including
12 CME-active ARs and 9 CME-quiet ARs.

For each event, based on the vector
magnetograms from Space-Weather Helioseismic and Magnetic Imager
(HMI; Scherrer et al. 2012) AR Patches (SHARP; Bobra et al. 2014) observed by \emph{Solar Dynamics
Observatory} (\emph{SDO}; Pesnell et al. 2012), we calculate $\Phi$$_\mathrm{AR}$ before the flare onset
by summing all pixels where vertical magnetic field $|$$B_{z}$$|$$>$100 G (Kazachenko et al. 2017). The magnetograms are re-mapped
using a cylindrical equal area (CEA) projection with a pixel size of
$\sim$0$\arcsec$.5 and presented as ($B_{r}$, $B_{\theta}$,
$B_{\phi}$) in heliocentric spherical coordinates corresponding to
($B_{z}$, -$B_{y}$, $B_{x}$) in heliographic coordinates (Sun 2013).
We have identified a ``flaring polarity inversion line" (FPIL)
mask to demarcate the core of an AR by using the method of Sun et al. (2015).
We first find the polarity inversion line (PIL) pixels from a smoothed vertical magnetic field
$B_{z}$, and dilate them with a circular kernel with a radius of 18 pixels (about 6.5 Mm; other radii are also used and the results are not affected).
Then we isolate flare ribbons by using the 1600 {\AA} image near the flare peak from the Atmospheric Imaging Assembly
(AIA; Lemen et al. 2012) on board the \emph{SDO} (above 700 DN $s^{-1}$ which is about 10 times the standard deviation above the mean of the quiet-Sun values), and dilate them with a large kernel having a radius of 20 pixels (about 7.2 Mm). Finally the intersection of
dilated PIL and ribbon areas is considered as the FPIL mask. The FPIL mask is similar to the strong-field, high-gradient PIL mask in Schrijver (2007), however, it only involves part of the PIL mask.

We calculate distributions of vertical electric current density $J_{z}$ and the mean characteristic
twist parameter $\alpha$$_\mathrm{FPIL}$ within the FPIL mask region.
We note that the FPIL mask region could be determined only after the flare occurrence and thus
$\alpha$$_\mathrm{FPIL}$ can not be used for CME forecasting. The energy release during a flare is generally believed to
originate from areas with high values of photospheric magnetic free energy density $\rho_\mathrm{free}$ in ARs, and thus we select the region with $\rho_\mathrm{free}$$>$4.0$\times$10$^{4}$ erg cm$^{-3}$
(HFED region; Chen \& Wang 2012) as a proxy for AR core region and calculate mean characteristic twist parameter $\alpha$$_\mathrm{HFED}$ and mean shear angle $\Psi$$_\mathrm{HFED}$ within
HFED region. Compared with $\alpha$$_\mathrm{FPIL}$, $\alpha$$_\mathrm{HFED}$ does not need ribbon information and is more suitable for CME forecasting.
Detailed formulas of the parameters are listed in Table 1.

\section{Statistical Results}

Figure 1 shows
four examples of two eruptive and two confined flares, which includes 1600 {\AA} images (left), photospheric magnetograms (middle) and derived vertical electric current density $J_{z}$ maps
(right). We can see that for two eruptive flares (X2.1 flare in AR 11283 and M6.5 in AR 12371) the positive
and negative currents have a coherent structure around the PIL (Figures 1(a)-(b)), indicating the presence of ``current ribbons" as in a
coherent flux rope. However, the two confined events (X3.1 in AR 12192 and M6.1 in AR 12222) exhibit disordered current distributions
and do not have any noticeable structure (Figures 1(c)-(d)). It also can been seen that the FPIL mask regions (orange and black contours) overlap the area of large currents,
implying that the FPIL mask corresponds to the AR core with the strongest magnetic non-potentiality. We estimate the errors of $J_{z}$ to be about 10 mA m$^{-2}$ based on
the noise level of transverse magnetic field ($\sim$ 100 G in Liu et al. 2012).

Based on the derived $J_{z}$ map, we then calculated the mean characteristic twist parameter $\alpha$$_\mathrm{FPIL}$ within the FPIL mask
region for the 106 flares. Figure 2(a) shows the scatter plot of
$\alpha$$_\mathrm{FPIL}$
versus $\Phi$$_\mathrm{AR}$. Blue (red) circles are the eruptive (confined)
flares. It needs to be noted that $\alpha$$_\mathrm{FPIL}$ is a signed parameter and in our study $\alpha$$_\mathrm{FPIL}$ means its absolute value.
It can be seen that the events with $\alpha$$_\mathrm{FPIL}$$<$0.07 Mm$^{-1}$ (black dotted line in Figure
2(a)) and $\Phi$$_\mathrm{AR}$$>$1.0$\times$$10^{23}$ Mx (right green dashed line in Figure
2(a)) are all confined and those with $\alpha$$_\mathrm{FPIL}$$\geq$0.07 Mm$^{-1}$ and $\Phi$$_\mathrm{AR}$$<$3.5$\times$$10^{22}$ Mx (left green dashed line in Figure
2(a)) are all eruptive. According to error propagation theory, we estimate the errors of $\alpha$$_\mathrm{FPIL}$
to be $10^{-4}$$-$$10^{-3}$ Mm$^{-1}$ (by considering the error of $J_{z}$ $\sim$ 10 mA m$^{-2}$). The errors are small and they do not affect
our results. The noise level of $B_{z}$ is on the order of 10 G (Liu et al. 2012),
and the errors in the calculation of $\Phi$$_\mathrm{AR}$ are estimated to be
about $10^{18}$$-$$10^{19}$ Mx. The errors are much smaller than $\Phi$$_\mathrm{AR}$ and thus are not considered. Figure 2(b) shows the scatter plot of
flare peak X-ray flux versus $\alpha$$_\mathrm{FPIL}$/$\Phi$$_\mathrm{AR}$. About 93\% (40 of 43) of eruptive events have
$\alpha$$_\mathrm{FPIL}$/$\Phi$$_\mathrm{AR}$$\geq$2.2$\times$$10^{-24}$ Mm$^{-1}$ Mx$^{-1}$, and $\sim$ 83\% (52 of 63) of confined flares have
$\alpha$$_\mathrm{FPIL}$/$\Phi$$_\mathrm{AR}$$<$2.2$\times$$10^{-24}$ Mm$^{-1}$ Mx$^{-1}$ (black dash-dotted line in Figure
2(b)). This shows that the new relative non-potential parameter $\alpha$$_\mathrm{FPIL}$/$\Phi$$_\mathrm{AR}$
is well able to distinguish the two populations of eruptive and confined flares. The statistical comparison between the two distributions in Figure 2(b) was done with a two-sample t-test, which shows a significant difference (P$<$0.001).
However, there is still a small overlap (14 of 106) by using our criterion and the exceptional events are mainly from
a CME-active AR 11302 and two CME-quiet ARs 11476 and 12268.

Figure 3 shows the maps of photospheric magnetic free energy density $\rho_\mathrm{free}$
and magnetic shear angle $\Psi$ for the four examples shown in Figure 1. Free magnetic energy is the amount of magnetic energy in
excess of the minimum energy attributed to the potential field, and magnetic shear is defined as the angle between the horizontal components of the observed
magnetic field and a modeled potential magnetic field based
on photospheric $B_{z}$ map. They are commonly used parameters in describing the magnetic complexity and
non-potentiality (Wang et al. 1994; Su et al. 2014). It can be seen that the maps of $\rho_\mathrm{free}$ and $\Psi$
exhibit similar distributions, with their large values around the PILs of ARs. The comparison of two eruptive flares (Figures 3(a)-(b))
with two confined ones (Figures 3(c)-(d)) shows that the mean values of $\rho_\mathrm{free}$ and $\Psi$ of eruptive flares are larger
than those of confined events. Similar to the appearance of $J_{z}$ maps (Figure 1), $\rho_\mathrm{free}$ and $\Psi$ of eruptive flares exhibit ribbon patterns, however, the $\rho_{free}$ and $\Psi$
maps of confined flares show disordered distributions.

We determine high-$\rho_\mathrm{free}$ areas (HFED region) as a proxy for AR core region and then calculate mean characteristic twist parameter $\alpha$$_\mathrm{HFED}$ and mean shear angle $\Psi$$_\mathrm{HFED}$ within the
HFED region. Figure 4 shows the calculation results for 43 eruptive and 63 confined flares.
It can be seen that the distributions of $\alpha$$_\mathrm{HFED}$ (Figure 4(a)) are similar to those of $\alpha$$_\mathrm{FPIL}$ (Figure 2(a)).
For $\alpha$$_\mathrm{HFED}$$<$0.1 Mm$^{-1}$ (black dotted line in Figure
4(a)), an overwhelming majority (about 93\%, 26 out of 28) of flares are confined. Almost all the
eruptive flares (41 out of 43) have $\alpha$$_\mathrm{HFED}$$\geq$0.1 Mm$^{-1}$. If we consider the relative parameter $\alpha$$_\mathrm{HFED}$/$\Phi$$_\mathrm{AR}$,
the differences between confined and eruptive cases are more evident (Figure 4(b)).
About 91\% (39 of 43) of eruptive flares have
$\alpha$$_\mathrm{HFED}$/$\Phi$$_\mathrm{AR}$$\geq$3.2$\times$$10^{-24}$ Mm$^{-1}$ Mx$^{-1}$, and $\sim$ 75\% (47 of 63) of confined events have
$\alpha$$_\mathrm{HFED}$/$\Phi$$_\mathrm{AR}$$<$3.2$\times$$10^{-24}$ Mm$^{-1}$ Mx$^{-1}$ (black dash-dotted line in Figure
4(b)). There is also a difference of mean shear angle $\Psi$$_\mathrm{HFED}$ within HFED region
between eruptive and confined flares (Figure 4(c)). The proportion of confined
flares is $\sim$91\% (29 of 32) corresponding
to $\Psi$$_\mathrm{HFED}$$<$60$^{\circ}$. Similarly, the relative parameter $\Psi$$_\mathrm{HFED}$/$\Phi$$_\mathrm{AR}$
can provide a good ability for distinguishing the eruptive and confined events (Figure 4(d)).
An overwhelming majority of eruptive flares have $\Psi$$_\mathrm{HFED}$/$\Phi$$_\mathrm{AR}$$\geq$1.0$\times$$10^{-21}$ degree Mx$^{-1}$
and most of confined flares show $\Psi$$_\mathrm{HFED}$/$\Phi$$_\mathrm{AR}$$<$1.0$\times$$10^{-21}$ degree Mx$^{-1}$.
The degree of overlap between eruptive and confined events for $\Psi$$_\mathrm{HFED}$/$\Phi$$_{AR}$
is a little higher than that for $\alpha$$_\mathrm{HFED}$/$\Phi$$_\mathrm{AR}$. The two-sample t-test was carried out for the two parameters in Figures 4(b) and (d), and showed a significant difference (P$<$0.001).

\section{Summary and Discussion}

In this study, we have analyzed the magnetic non-potentiality of ARs and the constraining effect of background fields for
106 flares $\geq$M1.0-class from 12 CME-active and 9 CME-quiet ARs. We proposed a new parameter $\alpha$$_\mathrm{FPIL}$/$\Phi$$_\mathrm{AR}$ that
is well able to distinguish ARs with the capability of producing eruptive flares.
About 93\% of eruptive events have the distributions of $\alpha$$_\mathrm{FPIL}$/$\Phi$$_\mathrm{AR}$$\geq$2.2$\times$$10^{-24}$ Mm$^{-1}$ Mx$^{-1}$,
and $\sim$ 83\% of confined flares have $\alpha$$_\mathrm{FPIL}$/$\Phi$$_\mathrm{AR}$$<$2.2$\times$$10^{-24}$ Mm$^{-1}$ Mx$^{-1}$.
Moreover, we select the areas of high photospheric magnetic free energy
density before the flare onset as a proxy for AR core region to calculate mean characteristic twist parameter $\alpha$$_\mathrm{HFED}$ and mean shear angle $\Psi$$_\mathrm{HFED}$.
The statistical results showed that the relative parameters $\alpha$$_\mathrm{HFED}$/$\Phi$$_\mathrm{AR}$ and $\Psi$$_\mathrm{HFED}$/$\Phi$$_\mathrm{AR}$ can also
provide a good ability for distinguishing the eruptive and confined flares. Overall, about 81\% (86 out of 106) of flare events can be classified into
the two populations of confined and eruptive flares using the parameter $\alpha$$_\mathrm{HFED}$/$\Phi$$_\mathrm{AR}$ (through determining a critical value).
The performance of $\Psi$$_\mathrm{HFED}$/$\Phi$$_\mathrm{AR}$ is not as good as that of $\alpha$$_\mathrm{HFED}$/$\Phi$$_\mathrm{AR}$, and about 74\% (78 out of 106) of flare events can be
distinguished.

Parameter $\alpha$ is the average characteristic twist of
the magnetic field lines around the PILs of an AR, and provide measures of the nonpotentiality of
AR core region (Leka \& Skumanich 1999; Benson et al. 2021). Previous studies have shown that the parameter of
magnetic twist plays an important role in discriminating between confined and
eruptive events and can be used to predict whether an X- or M-class flaring AR would produce a CME (Bobra \& Ilonidis 2016; Duan et al. 2019).
Magnetic twist is thought to be related to the driving force of electric current-driven
instabilities, such as the helical kink instability (Hood \& Priest 1979). An AR containing a highly twisted magnetic field
tends to produce an eruption when the twist exceeds a certain threshold. On the other hand,
according to the study of Li et al. (2020), $\Phi$$_\mathrm{AR}$ has a high positive correlation with the critical decay index height (related to the torus instability of a
magnetic flux rope; Kliem \& T{\"o}r{\"o}k 2006), implying that
$\Phi$$_\mathrm{AR}$ describes the strength of the background field confinement. Our statistical study reveals that the relative parameter $\alpha$/$\Phi$$_\mathrm{AR}$
has a better performance in distinguishing between the two types of
flares than only $\Phi$$_\mathrm{AR}$ or $\alpha$ does.
We suggest that the relative parameter $\alpha$/$\Phi$$_\mathrm{AR}$
indicates the balance between the upward force that drives the eruptions and the downward force that suppresses the eruptions.

Recently, several studies have shown the general trend that confined
events have a smaller reconnection flux fraction ($\Phi$$_\mathrm{ribbon}$/$\Phi$$_\mathrm{AR}$; defined by the ratio between the flux swept by the flare ribbons to the total AR flux)
compared with eruptive flares (Toriumi et al. 2017; Li et al. 2020; Kazachenko et al. 2021).
However, there is a considerable overlap between the flux fraction distributions of confined and eruptive flares.
Comparably, our parameters $\alpha$/$\Phi$$_\mathrm{AR}$ and $\alpha$$_\mathrm{HFED}$/$\Phi$$_\mathrm{AR}$
show more significant differences in distributions between the two populations of flares.
We hypothesize that $\Phi$$_\mathrm{ribbon}$ involves part of overlying background confining fields of ARs participating the flare reconnection at the late stage
of the flare. In comparison, $\alpha$$_\mathrm{FPIL}$ and $\alpha$$_\mathrm{HFED}$ only correspond to the pre-flare high-twist fields in an AR core region, which are strongly related to the
initial driving force of solar flares. Therefore, $\alpha$$_\mathrm{FPIL}$/$\Phi$$_\mathrm{AR}$ and $\alpha$$_\mathrm{HFED}$/$\Phi$$_\mathrm{AR}$ can provide a better ability for distinguishing the eruptive and confined
events than $\Phi$$_\mathrm{ribbon}$/$\Phi$$_\mathrm{AR}$.

The idea of the new parameter $\alpha$/$\Phi$$_\mathrm{AR}$ can be generalized to ``relative non-potentiality",
which refers to the ratio of magnetic flux (or other physical quantities) in a flux rope to that in the surrounding magnetic structures (Lin et al. 2021).
A larger relative non-potentiality indicates a higher probability for a flux rope to erupt (Toriumi et al. 2017; Thalmann et al. 2019; Gupta et al. 2021).
Recently, Lin et al. (2020) proposed a new relative non-potentiality parameter of the magnetic flux in the highly twisted region relative to its ambient
background fields, which demonstrated a moderate ability in discriminating between confined and
ejective events. It seems that our parameters $\alpha$/$\Phi$$_\mathrm{AR}$ shows a better classification performance than the parameter proposed in Lin et al. (2020). For instance, two events of X1.6 and X3.1-class flares in AR 12192 failed to
be correctly classified in Lin et al. (2020, 2021), however, they can be correctly classified in our study.
However, due to the difference of selected samples, a direct comparison can not be carried out.

In recent years, the importance of the confinement of flux ropes by overlying loops (and thus the importance of structural relativity) has been actively discussed in the theoretical and modeling studies. Leake et al. (2013) simulated a magnetic flux emergence into pre-existing dipole coronal field and found that it becomes a stable flux rope if the dipole coronal field is orientated to minimize magnetic reconnection. Using the same simulation results, Pariat et al. (2017) found that the ratio of the magnetic helicity of the current-carrying magnetic field to the total relative helicity diagnoses very clearly the eruptive potential of their parametric simulations. Toriumi \& Takasao (2017) conducted flux emergence models and showed that the confinement of the flux rope or the access to the outer space depends on the large-scale AR structures, which determines the CME eruption. The CME eruption is predicted by the SHARP parameters that characterize the ``relativity" between the flaring zone and overall AR area. Amari et al. (2018) suggested that the role of the magnetic cage is one deciding factor for the success or failure of CMEs.

Our findings imply that the relative measure of magnetic nonpotentiality within the AR core over the restriction of the background field largely controls whether a flare is eruptive or confined. However, it needs to be noted that there is still a small overlap (less than 20\%) between the two populations of confined and eruptive flares by using parameter $\alpha$/$\Phi$$_\mathrm{AR}$.
We suggest that other unknown mechanisms or intrinsic stochasticity may also play a role in governing CME production.

\acknowledgments {We are grateful to Dr. Jiangtao Su for useful discussions.
This work is supported by the B-type Strategic Priority
Program of the Chinese Academy of Sciences (XDB41000000), the National Key R\&D
Program of China (2019YFA0405000), the
National Natural Science Foundations of China (11773039, 11903050,
12073001, 11790304, 11873059 and 11790300), Key Programs of the
Chinese Academy of Sciences (QYZDJ-SSW-SLH050), the Youth Innovation
Promotion Association of CAS (2014043), Yunnan
Academician Workstation of Wang Jingxiu (No. 202005AF150025) and
NAOC Nebula Talents Program. X. D. Sun acknowledges the support by NSF award 1848250.
A. Q. Chen is supported by the
Strategic Priority Program on Space Science, Chinese Academy of
Sciences, Grant No. XDA15350203. \emph{SDO}
is a mission of NASA's Living With a Star Program.}

{}
\clearpage

\begin{table*}
\centering \caption{Parameters Used to Distinguish the Eruptive and Confined Flares\tablenotemark{1} \label{tab1}} \centering
\begin{tabular}{c c c c c c c} 
\hline\hline 
Parameters & Description & Unit &
Formula \\ 
\hline 
$\Phi$$_{AR}$ & Total unsigned flux & Mx & $\Phi$$_{AR}$=$\Sigma$$|$$B_{z}$$|$dA \\ J$_{z}$ & Mean vertical electric current density & mA m$^{-2}$ & J$_{z}$=$\frac{1}{N\mu}$$\Sigma$($\nabla$$\times$\textbf{B})$_{z}$ \tablenotemark{2} \\
$\alpha$ & Mean characteristic twist parameter & Mm$^{-1}$ & $\alpha$=$\frac{\mu\Sigma J_{z}B_{z}}{\Sigma B_{z}^{2}}$ \\ $\rho_{free}$ & Magnetic free energy density & erg cm$^{-3}$ & $\rho_{free}$=$\frac{1}{8\pi}$$|$$\textbf{B}$$_{obs}$-$\textbf{B}$$_{pot}$$|$$^{2}$\\
$\Psi$ & Mean shear angle & degree & $\Psi$=$\arccos$$\frac{\textbf{B}_{obs}\cdot\textbf{B}_{pot}}{|
 B_{obs}B_{pot}|}$\\
\\\hline
\end{tabular}
\tablenotetext{1}{Adapted from Chen \& Wang (2012) and Bobra et al. (2014).}
\tablenotetext{2}{$\mu$ is the magnetic permeability in vacuum (4$\pi$$\times$$10^{-3}$ G m $A^{-1}$).}
\end{table*}

\begin{figure}
\centering
\includegraphics
[bb=23 120 565 708,clip,angle=0,scale=0.78]{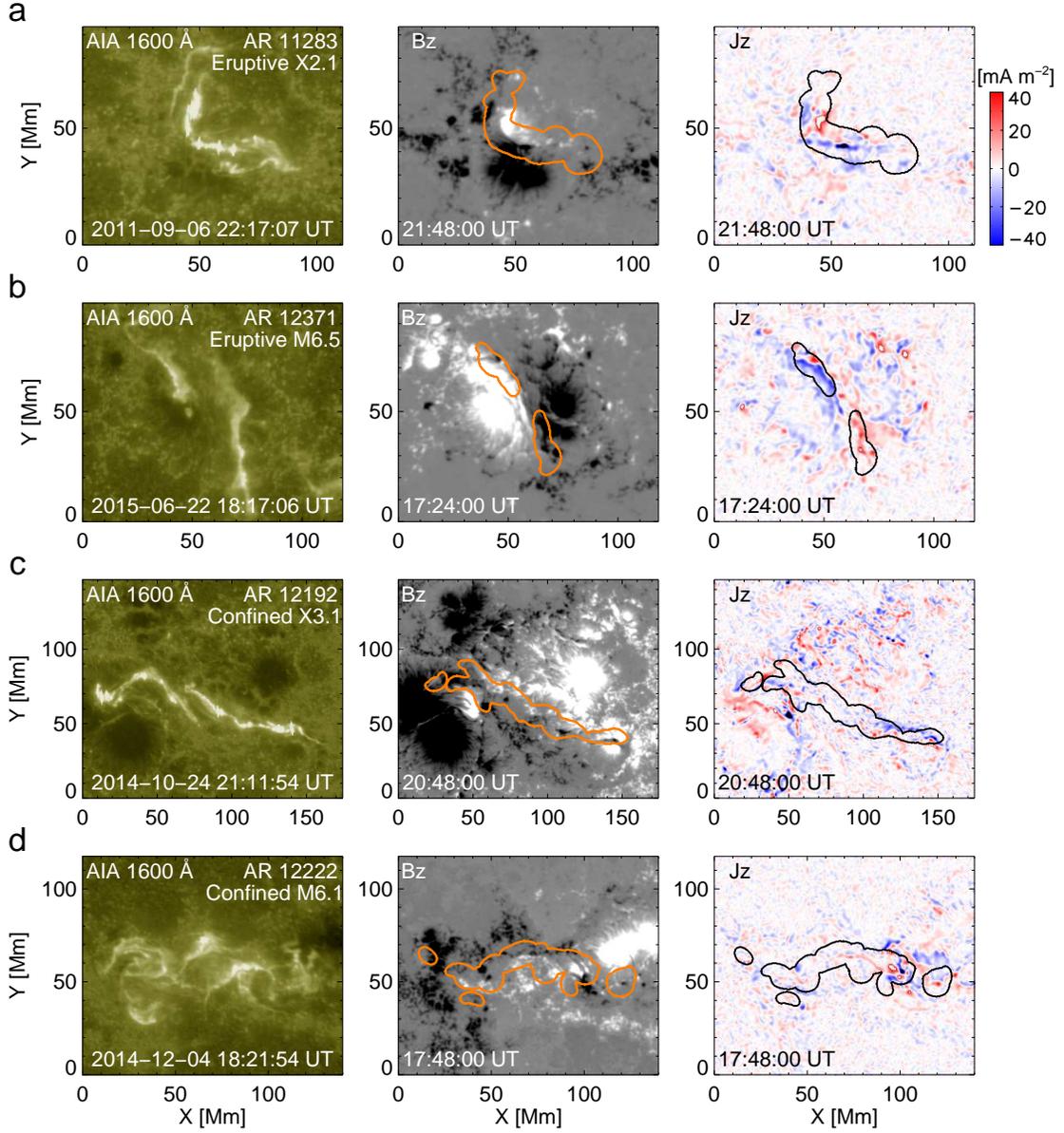}
\caption{Four examples of two eruptive and two confined flares showing SDO/AIA 1600 {\AA} images (left),
SDO/HMI photospheric magnetograms $B_{z}$ (middle) and derived vertical electric current density $J_{z}$ maps (right). From top to bottom: eruptive X2.1-class flare in AR 11283, eruptive M6.5-class flare in AR 12371,
confined X3.1-class flare in AR 12192 and confined M6.1-class flare in AR 12222. AIA 1600 {\AA} images were remapped with CEA projection. Orange and black contours outline
the FPIL mask regions within which the mean characteristic twist parameter ($\alpha$$_\mathrm{FPIL}$) in Figure 2 was calculated.
\label{fig1}}
\end{figure}
\clearpage

\begin{figure}
\centering
\includegraphics
[bb=14 321 531 505,clip,angle=0,scale=0.85]{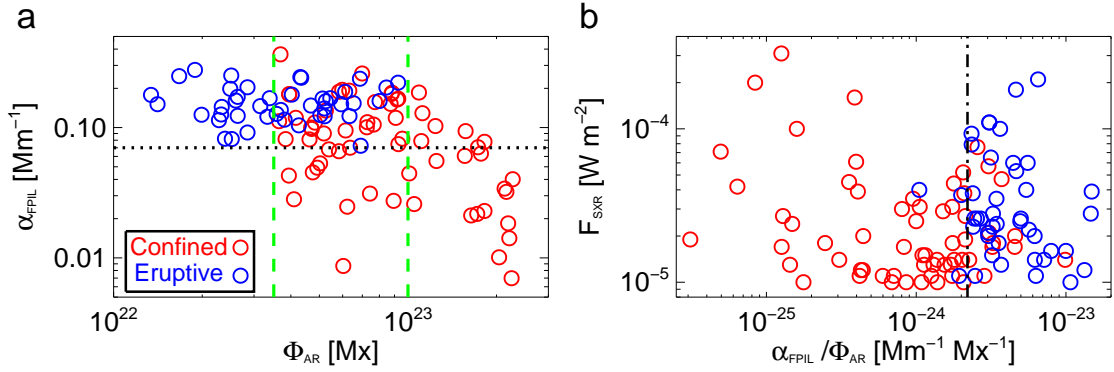}
\caption{Scatter plots of mean characteristic twist parameter $\alpha$$_\mathrm{FPIL}$ versus total unsigned magnetic flux $\Phi$$_\mathrm{AR}$ and
flare peak X-ray flux F$_\mathrm{SXR}$ versus $\alpha$$_\mathrm{FPIL}$/$\Phi$$_\mathrm{AR}$.
Blue (red) circles are eruptive (confined)
flares. Two vertical green lines in panel (a) correspond to $\Phi$$_\mathrm{AR}$ of
3.5$\times$$10^{22}$ Mx and 1.0$\times$$10^{23}$ Mx, respectively. The horizontal black line in panel (a) refers to
$\alpha$$_\mathrm{FPIL}$ of 0.07 Mm$^{-1}$. The vertical black line in panel (b) denotes $\alpha$$_\mathrm{FPIL}$/$\Phi$$_\mathrm{AR}$ of 2.2$\times$$10^{-24}$ Mm$^{-1}$ Mx$^{-1}$.
\label{fig2}}
\end{figure}
\clearpage

\begin{figure}
\centering
\includegraphics
[bb=14 3 580 820,clip,angle=0,scale=0.6]{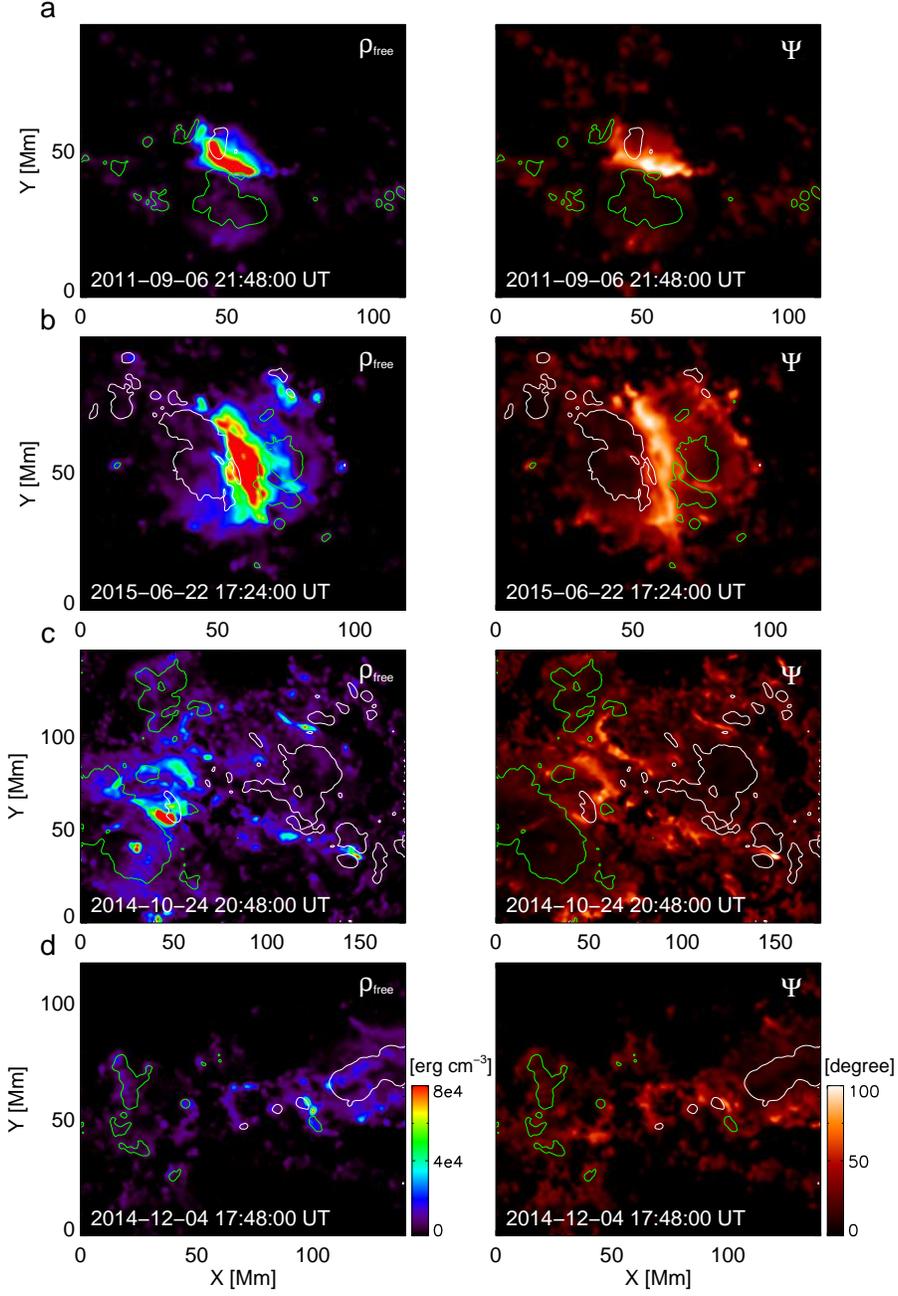}
\caption{Maps of photospheric free magnetic
energy density $\rho_\mathrm{free}$ and magnetic shear angle $\Psi$ of four examples (from top to bottom: eruptive X2.1-class flare in AR 11283, eruptive M6.5-class flare in AR 12371,
confined X3.1-class flare in AR 12192 and confined M6.1-class flare in AR 12222).
The white and green contours are the magnetic fields $B_{z}$
at $\pm$800 G levels. Mean characteristic twist parameter $\alpha$$_\mathrm{HFED}$
and mean shear angle $\Psi$$_\mathrm{HFED}$ in Figure 4 are calculated within the areas of $\rho_\mathrm{free}$$>$4.0$\times$10$^{4}$ erg cm$^{-3}$.
\label{fig3}}
\end{figure}
\clearpage

\begin{figure}
\centering
\includegraphics
[bb=30 242 531 587,clip,angle=0,scale=0.8]{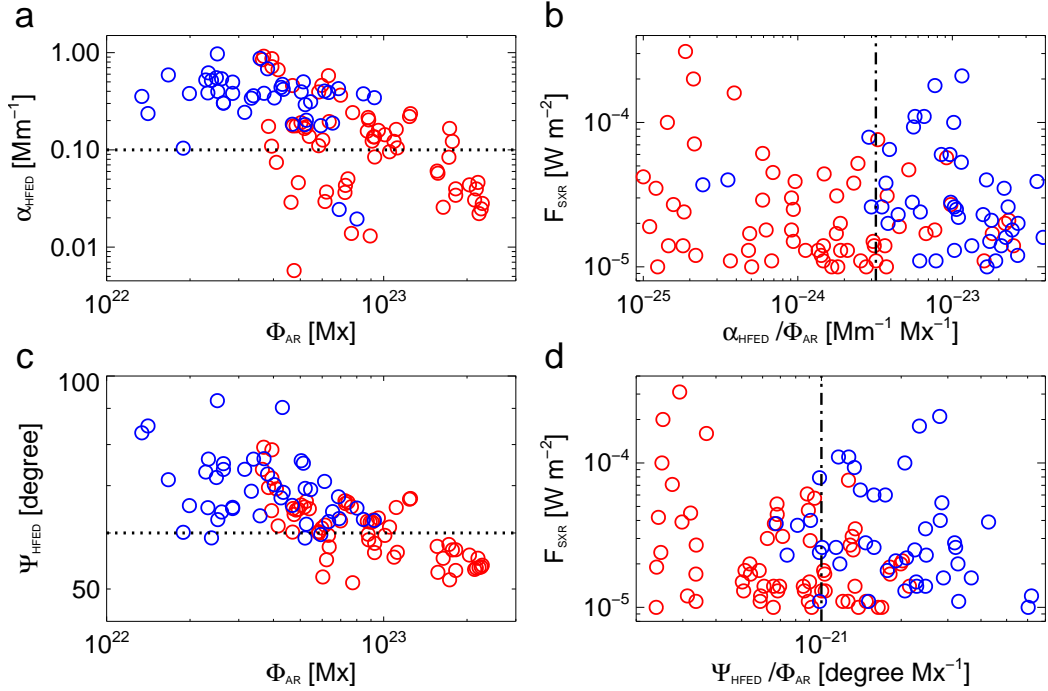} \caption{
Similar to Figure 2, but for mean characteristic twist parameter $\alpha$$_\mathrm{HFED}$
and mean shear angle $\Psi$$_\mathrm{HFED}$ within HFED region. Blue (red) circles are eruptive (confined)
flares. The horizontal dotted line in panel (a) corresponds to $\alpha$$_\mathrm{HFED}$ of 0.1 Mm$^{-1}$. The vertical dash-dotted line in panel (b) refers to
$\alpha$$_\mathrm{HFED}$/$\Phi$$_\mathrm{AR}$ of 3.2$\times$$10^{-24}$ Mm$^{-1}$ Mx$^{-1}$. The horizontal dotted line in panel (c) denotes $\Psi$$_\mathrm{HFED}$ of 60$^{\circ}$.
The vertical dash-dotted line in panel (d) corresponds to
$\Psi$$_\mathrm{HFED}$/$\Phi$$_\mathrm{AR}$ of 1.0$\times$$10^{-21}$ degree Mx$^{-1}$. \label{fig4}}
\end{figure}
\clearpage

\end{document}